\documentclass[singlecolumn, letterpaper, 10pt, prd, aps]{revtex4-1}
\usepackage[english]{babel}
\usepackage{graphicx}
\usepackage{textcomp}
\usepackage{amsmath}
\usepackage{amsfonts}
\usepackage{amssymb}
\usepackage{tikz}
\usetikzlibrary{arrows,positioning}
\usepackage{mathrsfs}
\usepackage{hyperref}
\usepackage{xcolor}
\def \ee{\end{equation}}
\def \be{\begin{equation}}
\def \eea{\end{eqnarray}}
\def \bea{\begin{eqnarray}}

\begin{document}

\title{On the Possibility of Non-Geodesic Motion of the Massless Spinning Top}
\author{Crist\'obal Armaza$^1$, Sergio A. Hojman$^{2,3,4}$, Benjamin Koch$^1$, and Nicol\'as Zalaquett$^1$}
\affiliation{$^1$Instituto de F\'isica, Pontificia Universidad Cat\'olica de Chile, Av. Vicu\~na Mackenna 4860, 782-0436 Macul, Santiago, Chile;\\
$^{2}$Departamento de Ciencias,
Facultad de Artes Liberales, Facultad de Ingenier\'{\i}a y Ciencias,\\
Universidad Adolfo Ib\'a\~nez, Santiago, Chile;\\
$^{3}$ Departamento
de F\'{\i}sica, Facultad de Ciencias, 
Universidad de Chile, Santiago, Chile;\\ 
$^{4}$ Centro de Recursos Educativos Avanzados,
CREA, Santiago, Chile}
\date{\today}

\begin{@twocolumnfalse}
\begin{abstract} 
The motion of spinning massless particles in gravitationally curved backgrounds
is revisited by considering new types of constraints. 
Those constraints guarantee
zero mass ($P_\mu P^\mu=0$)
and they allow for the possibility of trajectories which
are not simply null geodesics. To exemplify this previously unknown possibility,
the equations of motion are solved for radial motion in Schwarzschild background.
It is found that the particle experiences a spin-induced energy shift, which is
proportional to the Hawking temperature of the black hole background. 
\end{abstract}
\end{@twocolumnfalse}

\maketitle
\section{Introduction}

It is known that particles without internal structure travel on geodesics in curved spacetime \cite{Wein}.
It is further known that particles with internal rotational degrees of freedom, called spinning tops (STOPs),
can travel on modified (non-geodesic) trajectories in 
curved backgrounds. This fact has been continuously investigated
throughout almost a century \cite{Frenkel:1926zz,mat37,Weyss46,pap51,Corinaldesi:1951pb,tulc59,Pirani:1956tn,taub64,dixon64,vz69,Wald:1972sz,hanson74,hojman75,Barker:1975ae,Barducci:1976xq,hojman77,hojman78,Audretsch:1981wf,Audretsch:1981xn,Seitz:1986sc,Spinosa:1987sg,Spinosa:1987iu,Khriplovich:1997ni,Costa:2011zn,Costa:2012cy,Hojman:2012tt,Hojman2013,Zalaquett:2014eia,Costa:2014nta,Deriglazov:2014tsa,Deriglazov:2015zta,Armaza:2015eha}.
The degrees of freedom of the corresponding equations of motion have to be complemented by
additional conditions which involve the spin tensor $S^{\mu \nu}$ and either
the momentum $P_\mu$ or the velocity $U_\mu$.
This has been well studied for massive STOPs, where either the 
Frenkel-Mathisson-Pirani \cite{Frenkel:1926zz,mat37,Pirani:1956tn} conditions,
\be\label{FMP}
S^{\mu \nu}U_\nu=0,
\ee
or the Tulczyjew-Dixon \cite{tulc59,dixon64} conditions,
\be\label{Tul}
S^{\mu \nu}P_\nu=0,
\ee
are known. 
The latter are particularly attractive, since they imply the conservation of the invariant momentum squared, 
which is associated with mass.
Interestingly,
it is found that the deviations from usual geodesics increase with decreasing mass \cite{Hojman:2012tt,Armaza:2015eha}.
However, the transition to zero mass is not continuous.
For the description of massless STOPs,
both types of constraints (\ref{FMP} and \ref{Tul}) have been considered in 
a modified form.

In \cite{Bailyn:1977uj}, the Frenkel-Mathisson-Pirani type of constraint (\ref{FMP}) has been considered
for massless particles by imposing
\be\label{FMPm0}
S^{\mu \nu}U_\nu=a U^\mu , \quad \; P^\mu U_\mu=\frac{da}{d\tau}.
\ee
It was shown that the scalar $a$ is necessarily a constant.
It was further shown that if $U^\mu U_\mu \neq 0$ then necessarily $a=0$.
Moreover, if $U^\mu U_\mu =0$, then $a$ was chosen to be zero by ``initial condition''.
In \cite{Bailyn:1980}, the authors extended their discussion to the case of $a \neq 0$.
It has been shown that in this case, null geodesics
ensue without any further assumptions. The case $a=0$ allowes in principle
for trajectories which are not null geodesics.
This possibility of solutions which are no null geodesics has been considered
in \cite{dixon64,Bailyn:1980,Costa:2011zn}.

A Tulczyjew-Dixon type of constraint (involving momenta $P^\mu$ instead of velocities $U^\mu$) 
for massless particles has been investigated in \cite{mas75}.
By imposing
\be\label{Mashh}
S^{\mu \nu}P_\nu=0 , \quad \; P^\mu P_\mu=\frac{da}{d\tau},
\ee
the authors showed that a massless STOP  necessarily follows
null geodesics and that the spin is either parallel or antiparallel to the direction of motion.
This result has led to the common believe that a
zero mass ($P^\mu P_\mu=0$) necessarily implies motion on null geodesics, independently of
spinning degrees of freedom.
To the authors' knowledge, no relaxed version of the constraint (\ref{Mashh}), 
for instance
\be\label{Mashh2}
S^{\mu \nu}P_\nu=\alpha P^\mu , 
\ee
has been studied in the literature for massless STOPs.
This is surprising since it is actually the constraints of this type that
give a proper notion of conserved mass $P_\mu P^\mu=const$.

The aim of this paper is to revisit the scenario of 
describing massless  STOPs by considering constraints of the modified Tulczyjew-Dixon type (those involving momenta).
The paper is organized as follows.
In section \ref{secSTOP0} the physical quantities 
used to describe the motion of the STOP are introduced
and the corresponding equations of motion and the conserved quantities 
of the theory are presented.
Section  \ref{sec_Constr} presents 
 a set of possible constraints describing a massless STOP.
In section \ref{secsol}, a particular solution 
of the equations of motion for a massless STOP
subject to the new constraints is derived. This solution is shortly discussed
and surprising features are mentioned.
Conclusions are drawn in section \ref{secConcl}.

\section{Description of the STOP}
\label{secSTOP0}

\subsection{Dynamic Variables}
\label{secSTOP}
The description of the STOP will follow
the definitions and notations given in \cite{hojman75,hojman78}.
The position of a relativistic top is denoted by a
set of four coordinates $x^\mu$, while its orientation is defined by an
orthonormal tetrad ${e_{a}}^{\mu}(\tau)$
dependent of the particle's word line. 
A gravitational field is described in terms of the metric field $g_{\mu \nu}$. The tetrad vectors satisfy $g_{\mu \nu} \
{e_{a}}^{\mu} \ {e_{{b}}}^{\nu} \ \equiv\ \eta_{a
b}$, with $\eta_{a b} \ \equiv \ \ \mbox{diag}\ (+1,
-1, -1, -1)=\eta^{a b}$, they have therefore six components
which are independent of the metric. The velocity vector $U^\mu$ is defined in
terms of an arbitrary parameter $\tau$ by
\begin{equation}
U^\mu\equiv \frac{d x^\mu}{d \tau}\, . \label{vel}
\end{equation}
The antisymmetric angular velocity tensor $\sigma^{\mu \nu}$ is
\begin{equation}
\sigma^{\mu \nu}\ \equiv \eta^{a b}
{e_{a}}^{\mu}\frac{D{e_{b}}}{D \tau} ^{\nu}\ = \ -
\ \sigma^{\nu \mu},\label{sigma}
\end{equation}
where the covariant derivative $D{e_{b}}^{\nu}/D\tau $ is
defined in terms of the Christoffel symbols ${\Gamma^{\nu}}_{\rho
\alpha}$, as usual, by
\begin{equation}
\frac{D{e_{b}}^\nu}{D \tau} \ \equiv \
\frac{d{e_{b}}^\nu}{d \tau}\ + {\Gamma^{\nu}}_{\rho \alpha} \
{e_{b}}^{\rho}\ U^\alpha\, . \label{covder}
\end{equation}
General covariance is 
achieved most elegantly and unambiguously at the level of the
Lagrangian formulation \cite{hojman75} due to the fact that only
first derivatives of the dynamical variables are used in its
construction.
A possible Lagrangian is constructed as
an arbitrary function of four invariants $a_1\equiv U^\mu U_\mu,\
a_2\equiv \sigma^{\nu \mu} \sigma_{\mu \nu},\ a_3\equiv U_\alpha \sigma^{\alpha \beta}
\sigma_{\beta \gamma} U^\gamma$ and $a_4\equiv \sigma_{\alpha \beta} \sigma^{\beta \lambda } \sigma_{\lambda \rho} \sigma^{\rho \alpha}$. 
Let $L (a_1, a_2, a_3, a_4)$ represent a generic Lagrangian in terms of these scalars. 
The conjugated momentum vector $P_\mu$ and antisymmetric spin tensor
$S_{\mu \nu}$ are defined by
\begin{equation}
P_\mu \equiv -\frac{\partial L}{\partial U^\mu},\qquad
S_{\mu \nu} \equiv -\frac{\partial L}{\partial \sigma^{\mu \nu}} = -
S_{\nu \mu}.
\end{equation}
These conjugated momenta are 
\begin{align}
P^\mu&=-2 U^\mu L_1 - 2 \sigma^{\mu\alpha}\sigma_{\alpha\lambda} U^\lambda L_3  ,\label{PDef}\\
S^{\mu\nu}&=-4\sigma^{\nu\mu}L_2 -2(U^\mu \sigma^{\nu\lambda}U_\lambda-U^\nu \sigma^{\mu\lambda}U_\lambda)L_3 - 8  \sigma^{\nu \lambda } \sigma_{\lambda \rho} \sigma^{\rho \mu} L_4,\label{SDef}
\end{align}
where $L_i\equiv \frac{\partial L}{\partial a_i}$. 
In order to shorten notation of the following discussions we define
\begin{equation}
  \label{defV}
  V^\mu\equiv S^{\mu\nu}P_\nu.
\end{equation}
 An other important element of the following discussions is
the Pauli-Luba\'nski pseudovector, which is defined by 
\begin{equation}
  \label{plDef}
  W^\mu\equiv {S^*}^{\mu\nu}P_\nu = \frac{1}{2}\epsilon^{\mu\alpha\beta\nu}S_{\alpha\beta}P_\nu,
\end{equation}
where $\epsilon^{0123}=+(\det (g_{\mu\nu}))^{-1/2}.$
A contraction of this vector with itself gives the second Casimir invariant of the group (along with $P^\mu P_\mu$)
\begin{equation}
  \label{casPL}
  W^\mu W_\mu =V_\mu V^\mu -\frac{1}{2}P^\mu P_\mu S^{\alpha\beta}S_{\alpha\beta}.
\end{equation}

\subsection{Equations and Constants of Motion}
\label{seceom}

A large part of the structure of the equations of motion for 
a STOP can be obtained independently of the choice of particular constraints.
Those equations are, however, not sufficient to determine the 
solution of the system.
Therefore, in some cases, constraints are used from the start,
either by imposing the constraints or by choosing a Lagrangian which 
implies the constraints.
The subtlety of defining a massless
STOP is due to the different possibilities of choosing the constraints,
whereas the equations of motion are the same,
independent of those subtleties.
Those equation are well known in the literature namely
\begin{equation}
 \frac{D P^\mu}{D\tau}=-\frac{1}{2}{R^\mu}_{\nu\alpha\beta}U^\nu S^{\alpha\beta}
\label{momentummotion}
\end{equation}
and
\begin{equation}
\frac{D S^{\mu \nu}}{D\tau}=S^{\mu
\lambda}{\sigma_\lambda}^\nu-\sigma^{\mu
\lambda}{S_\lambda}^\nu=P^\mu U^\nu-U^\mu P^\nu. \label{spinmotion}
\end{equation}
In Appendix \ref{EOMfromL}, it is shown in more detail how
those equations arise from a Lagrangian formulation 
\cite{hojman75, hojman78}.
Regardless of the form of the Lagrangian, the following quantities are constant of motion for any given metric:
\begin{equation}
J^2\equiv \frac{1}{2} S^{\mu \nu}S_{\mu \nu}\,  \label{spin}
\end{equation}
and
\begin{equation}
S^4 \equiv S^{\mu\alpha}S_{\alpha\beta}S^{\beta\gamma}S_{\gamma\mu}.
\end{equation}
The fact that $J^{2}$ is constant can be checked by taking the time derivative and replacing the equation of motion for ${S^\mu}_\nu$ giving
\begin{equation}
\dot{J^2}=2\mathring{S^{\mu\nu}}S_{\mu\nu}=2(S^{\mu\lambda}{\sigma_\lambda}^\nu-\sigma^{\mu\lambda}{S_\lambda}^\nu )S_{\mu\nu}=-4{S^{\mu}}_\lambda{\sigma^\lambda}_\nu {S^\nu}_{\mu}=0.
\end{equation}
The last step is due to the fact that, upon using the antisymmetry of $S^{\mu\nu}$ and $\sigma^{\mu\nu}$, we get ${S^{\mu}}_\lambda{\sigma^\lambda}_\nu {S^\nu}_{\mu} = - {S^{\mu}}_\lambda{\sigma^\lambda}_\nu {S^\nu}_{\mu}$. 
The same argument applies to $S^4$. 
In the massive case it was shown in~\cite{Hojman2013} that 
\begin{equation}
m^2\equiv P^\mu P_\mu \label{mass}
\end{equation}
is also a constant of motion 
if one uses a Tulczyjew-Dixon type of constraint (\ref{Tul}). 
In the massless case this demonstration will have to wait until 
the constraints are presented.
Finally, a conserved quantity $C_{\xi}$ given by
\begin{equation}
C_{\xi}\equiv P^\mu \xi_\mu -\frac{1}{2}S^{\mu \nu}\xi_{\mu;\nu}, \label{ckilling}
\end{equation}
can be associated to any Killing vector $\xi_\mu$ of the metric
\begin{equation}
\xi_{\mu;\nu} + \xi_{\nu;\mu}=0. \label{killingeq}
\end{equation}
This can shown straight forwardly by differentiating the conserved quantity \cite{hojman75} and using the equations of motion as well as identities of the Riemann tensor.
It can also be shown by using the Noether theorem.
The formal derivation of (\ref{ckilling}) from the Noether theorem
is given in Appendix \ref{SecNoether}.
\section{Possible Constraints for Massless Particles}
\label{sec_Constr}

As the momenta $P^\mu$ along with the spin tensor $S^{\mu \nu}$ add up to 10 degrees of freedom, 
one needs to implement conditions 
in order to reduce the degrees to those that correspond to a moving rotating particle. 
For a massive particle one would like to have three 
rotational degrees of freedom and four degrees of freedom associated to
to displacements in spacetime. Thus, a proper constraint for massive particles should reduce $10-7=3$ degrees of freedom.
However, for massless particles there is no rest frame associated to the motion of the particle, which means
that one would like to have only three degrees of freedom associated to displacements and three rotational degrees of freedom.
Thus, in a superficial counting, a proper constraint for massless particles would have to reduce $10-6=4$ degrees of freedom.
The aim in this section is to find constraints
that are consistent with the description of massless STOPs $P_\mu P^\mu=0$.

\subsection{Example, ``Pauli-Luba\'nski'' Constraint}

The Pauli-Luba\'nski pseudovector (\ref{plDef}) is usually 
identified with the helicity, therefore it is natural to consider the constraint
\be\label{PLconstr}
W^\mu=\lambda P^\mu |_{\lambda \neq 0},
\ee
where ``$\neq 0$'' symbolizes finite and non-zero. 
Let us analyze the implications of this constraint, before considering
other possibilities.
From the antisymmetry of ${S^\mu}_\nu$ one sees that the constraint
(\ref{PLconstr}) implies $P^2=0=W^2$.
Inserting this into equation
(\ref{casPL}) gives $V^2=0$.
Since $V^2=P^2=0$ and since further $V_\mu P^\mu=0$, 
it follows that
 $V^\mu=\alpha P^\mu$ (see Appendix \ref{Atheorem}).
The constant $\alpha$ can either be zero or non-zero,
those two scenarios have to be discussed separately.
\begin{itemize}
\item[${\mathcal{A}}$)]
If $\alpha=0$: $\Rightarrow$ $V^\mu = 0$,
which implies due to the spin-relations given in Appendix \ref{SecJ2rel}, that$S^*S=0$ and $J^2=\lambda^2$.
This completes scenario ${\mathcal{A}}$, which can be
summarized by the relations
\be\label{scenA}
{\mathcal{A}}: \left\{ P^2=0,\;W^2=0,\;V^2=0,\; W^\mu=\lambda P^\mu,\; V^\mu=0,\; S^*S=0,\; J^2=\lambda^2\right\}.
\ee
\item[${\mathcal{B}}$)] If $\alpha\neq0$: 
 From $W^2=0$, $V^2=0$, and  $W_\mu V^\mu= P^\nu S^*_{\mu \nu}V^\mu=\alpha  P^\nu S^*_{\mu \nu}P^\mu=0$ follows 
that $W^\mu=\gamma V^\mu$ (see Appendix \ref{Atheorem}).
With the spin-relations given in Appendix \ref{SecJ2rel}, this implies
$ S^*S\propto \alpha \lambda$ and $J^2=\lambda^2-\alpha^2$.
This completes scenario ${\mathcal{B}}$ which can be summarized
by the relations
\be\label{scenB}
{\mathcal{B}}:\left\{ P^2=0,\;W^2=0,\;V^2=0,\; W^\mu=\lambda P^\mu,\; V^\mu=\alpha P^\mu,\; W^\mu=\gamma V^\mu,\; S^*S\propto\alpha \lambda,\; J^2=\lambda^2-\alpha^2\right\}.
\ee
\end{itemize}
Thus, the constraint (\ref{PLconstr}) is consistent with $P^2=0$ and
it allows for the scenarios ${\mathcal{A}}$ and ${\mathcal{B}}$.
Clearly ${\mathcal A}$ looks similar to ${\mathcal B}$ for $\alpha=0$, however, the
two cases have to be treated separately, 
since the proof of  ${\mathcal B}$ relies on $\alpha\neq0$.
\subsection{Other Constraints}

Equation (\ref{PLconstr}) is not the only constraint
that could be consistent with $P_\mu P^\mu=0$.
For example, one can consider 
other constraints involving $W^\mu$, $V^\mu$, or $P^\mu$.
The simplest candidates for this are either of the squared type $A_\mu A^\mu=0$
or of the parallel type $A^\mu=a B^\mu$.
There are in total six constraints of those types involving $W^\mu$, $V^\mu$, or $P^\mu$, namely
\be\label{Cpossible}
W^\mu=\lambda P^\mu |_{\lambda \neq 0}, \quad{\mbox{or}}\; \;
V^\mu=\alpha P^\mu |_{\alpha \neq 0}, \quad{\mbox{or}}\;\;
W^\mu=\gamma V^\mu |_{\gamma \neq 0} , \quad{\mbox{or}}\;\;
P_\mu P^\mu=0, \quad{\mbox{or}}\;\;
W_\mu W^\mu=0, \quad{\mbox{or}}\;\;
V_\mu V^\mu=0,
\ee
where the first possibility in this list has been in  the previous subsection.
Since the squared type constraints
give just one single algebraic relation, they
are not sufficient to consistently reduce the degrees of freedom and simultaneously imply $P^2=0$.
Therefore, one also has to consider combinations of (\ref{Cpossible}) involving two constraints.
The constraints (\ref{Cpossible}) are covered by discussing six cases
and the possible combinations of (\ref{Cpossible}) are covered by discussing fifteen cases,
which sums up to the discussion of twenty-one cases.
Surprisingly, the outcome of those twenty-one cases, is
described by the scenarios ${\mathcal{A}}$ and ${\mathcal{B}}$
given in (\ref{scenA}) and (\ref{scenB}).
In Table \ref{tabcond}, it is shown which conditions imply scenario (\ref{scenA}), or (\ref{scenB}).
Further cases for which the initial constraints are insufficient to reduce the degrees of freedom
are labeled by ``0''.
The cases where a combination of initial constraints is {\it redundant} since already
one of the two constraints is sufficient to obtain  ${\mathcal A}$ or  ${\mathcal B}$
are labeled by ``$R$'' (redundant).
Since this table of combinations of initial constraints is obviously symmetric,
 only the upper half of the entries is shown.
\begin{table}[htp]
\caption{Possible outcomes when using two initial constraints. Entries with ``$R$'' mean that imposing both constraints is redundant, since
already one of the constraints would allow to obtain ${\mathcal A}$ or ${\mathcal B}$.
Entries with ``-'' indicate that the constraint is not sufficient to reduce the degrees of freedom and derive ${\mathcal A}$ or ${\mathcal B}$.
When a constant ($\alpha,\; \lambda,\;\gamma$) is explicit in a constraint it is assumed
that it is finite. Note that the diagonal entries of the table are actually just one single constraint.
}
\label{tabcond}
\begin{center}
\begin{tabular}{|c||c|c|c|c|c|c|}
\hline
 & $W^\mu=\lambda P^\mu$&$V^\mu=\alpha P^\mu$&$W^\mu=\gamma V^\mu$  &$P_\mu P^\mu=0$ & $W_\mu W^\mu=0$ &$V_\mu V^\mu=0$\\
\hline
\hline
&  & & & & &\hskip 4cm{}\\
 $W^\mu=\lambda P^\mu$ & ${\mathcal A}$ $\&$ ${\mathcal B}$& ${\mathcal B}_R$ & ${\mathcal B}_R$  &${\mathcal A}_R$ $\&$ ${\mathcal B}_R$  &${\mathcal A}_R$ $\&$ ${\mathcal B}_R$ &${\mathcal A}_R$ $\&$ ${\mathcal B}_R$  \\
&  & &  & &  &\hskip 4cm{}\\
\hline
&  & &  & & & \hskip 4cm{}\\
$V^\mu=\alpha P^\mu$&  & ${\mathcal B}$&    ${\mathcal B}_R$ &  ${\mathcal B}_R$&    ${\mathcal B}_R$ &   ${\mathcal B}_R$\\
&  & &  & & &\hskip 4cm{}\\
\hline
&  & &  & & & \hskip 4cm{}\\
$W^\mu=\gamma V^\mu$& & & ?? &  ${\mathcal B}_{R}$ &   ${\mathcal B}_{R}$ & ${\mathcal B}_{R}$ \\
&  & &  & & &\hskip 4cm{}\\
\hline
&  & &  & & &\hskip 4cm{}\\
$P_\mu P^\mu=0$ & & & & - & ${\mathcal A}$ $\&$ ${\mathcal B}$& ${\mathcal A}$ $\&$ ${\mathcal B}$\\
&  & &  & & & \hskip 4cm{}\\
\hline
&  & &  & & &\hskip 4cm{}\\
$W_\mu W^\mu=0$ & & & &  & -&${\mathcal A}$ $\&$ ${\mathcal B}$\\
&  & &  & & &\hskip 4cm{}\\
\hline
&  & &  & & &\hskip 4cm{}\\
$V_\mu V^\mu=0$ & & & &  & & -\\
&  & &  & & &\hskip 4cm{}\\
\hline
\end{tabular}
\end{center}
\label{default}
\end{table}
Please note that assuming $\alpha=0$ corresponds to the Tulczyjew constraint,
for which it is known that the limit $m\rightarrow 0$ is ill-defined.
Note further that the tensor ${C^\mu}_\nu=\frac{1}{2}\epsilon^{\mu \alpha \beta \gamma} S_{\alpha \beta} (S^{-1})_{\gamma \nu}$, 
appearing for the constraint $W^\mu=\gamma V^\mu$ seems not to be necessarily antisymmetric.
However, due to the lack of a proof, we left the corresponding entry in 
the table \ref{tabcond} with question marks.
Apart from the systematic study of possible constraints, one can find
physical arguments for certain initial constraints.
For example, a condition on the Casimir invariant $W_\mu W^\mu$ for the massless case can be obtained
from demanding the existence of finite dimensional representations of the Poincar\'e group:
the little group of symmetries representing the rotations of a massless particle in flat spacetime represent rotations and translations in 2D. 
This group is non-compact. If we demand the group to be compact 
(in order for a quantum theory that has finite dimensional representations to be viable) we have to accept that the translation operators
of this little group are null.
This would be an argument to use $W^\mu W_\mu=0$ as initial constraint, however, if this relation turns out to 
be a result of a different constraint, or combination of constraints, those are equally valid.

The final result of the analysis given in this subsection can be summarized as follows.
Apart from the
constraint (\ref{PLconstr}) one can construct at least three other constraints which give
exactly the same result (without redundance). Those other constraints are:
 ($P^2=0$ with $W^2=0$), ($P^2=0$ with $V^2=0$), and ($V^2=0$ with $W^2=0$).
In addition to those three cases there is the case ($V^\mu=\alpha P^\mu$),
which also agrees with (\ref{PLconstr}) in scenario ${\mathcal B}$.

\subsection{Constancy of Constraints and of $P^2$}
We now turn to the issue of proving the constancy of $P^2$ along the trajectory. 
The main difference of the constraints presented can be attributed to the use of $\alpha\neq 0$ and/or $\lambda\neq 0$.
Let us consider the following generic constraint
\begin{equation}
  \label{demp21}
  M^{\mu\nu}P_{\nu}=\epsilon P^{\mu} ,
\end{equation}
Where $M^{\mu \nu}$ is an antisymmetric tensor. Covariant differentiation of the constraint along the line gives
\begin{equation}
  \label{demp22}
  \mathring{M^{\mu\nu}}P_{\nu}+M^{\mu\nu}\mathring{P_{\nu}}=\dot{\epsilon} P^{\mu}+\epsilon \mathring{P^{\mu}}.
\end{equation}
Contracting \eqref{demp22} with $P_\mu$ gives
\begin{equation}
  \label{demp23}
   P_\mu M^{\mu\nu}\mathring{P_{\nu}}=\dot{\epsilon} P^{\mu}P_\mu+\epsilon \mathring{P^{\mu}}P_\mu,
\end{equation}
and contracting \eqref{demp21} with $\mathring{P_\mu}$ gives
\begin{equation}
  \label{demp24}
  \mathring{P_\mu}M^{\mu\nu}P_{\nu}=\epsilon P^{\mu}\mathring{P_\mu} .
\end{equation}
Then adding \eqref{demp23} and \eqref{demp24} considering the antisymmetry of $M^{\mu \nu}$ gives
\begin{equation}
  \label{demp25}
  \frac{d}{d\tau}(P^{\mu}P_\mu)=-\frac{ \dot{\epsilon}}{\epsilon}P^{\mu}P_\mu.
\end{equation}
This implies that if $P^\mu P_\mu=0$ at some instant $\tau=\tau_0$,
then $P^\mu P_\mu=0$ along the whole trajectory, provided that $\epsilon \neq 0$.
Further in the case of $\alpha=0$ and $\lambda\neq 0$, one can see from \eqref{alpha} that 
$S^{*\alpha\beta}S_{\alpha\beta}=0$. Covariant derivation of this scalar gives
\begin{equation}
  \label{pfafder}
  \epsilon^{\mu\nu\alpha\beta}\mathring{S_{\mu\nu}}S_{\alpha\beta}=2S^{*\mu\nu}(P_\mu U_\nu-P_\nu U_\mu)=-4\lambda P^\nu U_\nu=0.
\end{equation}
Since $\lambda\neq 0$ one gets
\begin{equation}
  \label{pu0}
  P^\nu U_\nu=0.
\end{equation}
This same result can easily be obtained considering $\alpha \neq 0$ and differentiating $J^2=\frac{1}{2}S^{\mu\nu}S_{\mu\nu}$ as this is a constant 
of motion. This means that, \eqref{pu0} holds
for any of the presented constraints. 
One is tempted to say that $P^{\mu}\propto u^\mu$ but one does not really 
know whether $u^\mu$ can be spacelike in some situations.
Now, considering that $\mathring{S^{*\alpha\beta}}=\frac{1}{2}\epsilon^{\alpha \beta \mu \nu}\mathring{S_{\mu\nu}}$, one can see that
when $M^{\mu \nu}=S^{\mu \nu}$ or $M^{\mu \nu}=S^{*\mu \nu}$, and using Eqs. (\ref{spinmotion}) and \eqref{pu0} that
\begin{equation}
  \label{MPp0}
  \mathring{M^{\mu\nu}}P_\nu=0.
\end{equation}
Using this and contracting \eqref{demp22} with $\mathring{P_\mu}$ gives
\begin{equation}
  \label{parappuntop}
  \epsilon \mathring{P_\mu} \mathring{P^\mu}=0,
\end{equation}
where in each case there is one $\epsilon\neq 0$ ($\epsilon=\alpha$ or $\epsilon=\lambda$). So, using the 
result of Appendix \ref{Atheorem}
for
 $\mathring{P_\mu} P^\mu=0$, $P_\mu P^\mu=0$ and $\mathring{P_\mu} \mathring{P^\mu}=0$
one concludes that
\begin{equation}
  \label{PesKP}
  \mathring{P^\mu}= \kappa P^\mu,
\end{equation}
where $\kappa$ is a scalar.
Replacing \eqref{PesKP} in \eqref{demp22} gives $\dot{\epsilon}=0$ turning $\epsilon$ in a new constant of motion.

\section{A First Solution for Schwarzschild-RN-(A)dS Background}
\label{secsol}
\subsection{Setting the Stage}

In order to exemplify the findings made in section \ref{sec_Constr},
the equations for a massless spinning top will be solved
for the case of a special trajectory on a generic static spherically symmetric gravitational background. 
This solution will be possible for both scenarios (\ref{scenA} and \ref{scenB}).
The background metric for this scenario is
\begin{equation}
  \label{metricSSS}
  ds^2=g(r)dt^2 - \frac{c^2}{g(r)}dr^2-r^2 d\theta^2 -r^2 \sin(\theta) d\phi^2 .
\end{equation}
In order to see whether a radial solution exists 
(which is not guaranteed when spin comes into play),
one can consider the following initial conditions for a trajectory: 
$\theta=\frac{\pi}{2}$, $\phi=0$, $P^\theta=0$ and $P^\phi=0$. 
The equations should then 
give rise to $\dot{\theta}=0$, $\dot{P^\theta}=0$, $\dot{\phi}=0$ and $\dot{P^\phi}=0$.
The Killing vectors associated to the metric (\ref{metricSSS}) 
allow to write the following constants of motion:
\bea
  \label{HConst}
  E&=&g P^t-\frac{1}{2} g' S^{t r},\\
    \label{jConst}
  j&=&-r S^{r \phi },\\
    \label{C3}
  C_3&=&-r^2 S^{\theta  \phi },\\
    \label{C4}
  C_4&=&-r S^{r \theta }.
\eea
The metric-blind constants are
\begin{equation}
  \label{PPCompleta}
  P^{\mu}P_{\mu}=-\frac{c^2 \left(P^r\right)^2}{g}+g \left(P^t\right)^2
\end{equation}
and
\begin{equation}
  \label{SSCompleta}
  \frac{1}{2}S^{\mu\nu}S_{\mu\nu}=\frac{g \left(r^4 \left(S^{\theta  \phi }\right)^2-c^2 \left(S^{t r}\right)^2\right)+c^2 r^2 \left(\left(S^{r \theta }\right)^2+\left(S^{r \phi }\right)^2\right)-g^2 r^2 \left(\left(S^{t \theta }\right)^2+\left(S^{t \phi }\right)^2\right)}{g}.
\end{equation} 
The equations of motion for $P^\mu$ are
\bea
\label{P1}
 0& = & \dot{P^t}+\frac{\dot{t} P^r g'}{2 g}+\frac{\dot{r} P^t g'}{2 g}-\frac{r \dot{\theta } S^{t \theta } g'}{2 c^2}-\frac{r \dot{\phi } S^{t \phi } g'}{2 c^2}-\frac{\dot{r} S^{t r} g''}{2 g} ,\\
\label{P2}
 0 &= &  \dot{P^r}+\frac{g \dot{t} P^t g'}{2 c^2}-\frac{r \dot{\theta } S^{r \theta } g'}{2 c^2}-\frac{r \dot{\phi } S^{r \phi } g'}{2 c^2}-\frac{g \dot{t} S^{t r} g''}{2 c^2}-\frac{\dot{r} P^r g'}{2 g} , \\
\label{P3}
  0 &= &  \dot{P^{\theta }}+\frac{\dot{\theta } P^r}{r}+\dot{\phi } S^{\theta  \phi }+\frac{\dot{r} S^{r \theta } g'}{2 g r}-\frac{g \dot{\phi } S^{\theta  \phi }}{c^2}-\frac{g \dot{t} S^{t \theta } g'}{2 c^2 r} , \\
\label{P4}
  0 &= &  \dot{P^{\phi }}+\frac{\dot{\phi } P^r}{r}+\frac{g \dot{\theta } S^{\theta  \phi }}{c^2}-\dot{\theta } S^{\theta  \phi }+\frac{\dot{r} S^{r \phi } g'}{2 g r}-\frac{g \dot{t} S^{t \phi } g'}{2 c^2 r} . 
\eea
The equations for $S^{\mu \nu}$ are 
\bea
\label{S1}
 0 & = &\dot{S^{t r}}+\dot{t} P^r-\dot{r} P^t-\frac{g r \dot{\theta } S^{t \theta }}{c^2}-\frac{g r \dot{\phi } S^{t \phi }}{c^2}, \\
\label{S2}
 0 &= &\dot{S^{t \theta }}-\dot{\theta } P^t+\frac{\dot{\theta } S^{t r}}{r}+\frac{\dot{r} S^{t \theta }}{r}+\frac{\dot{t} S^{r \theta } g'}{2 g}+\frac{\dot{r} S^{t \theta } g'}{2 g} , \\
\label{S3}
 0 &= &\dot{S^{t \phi }}-\dot{\phi } P^t+\frac{\dot{\phi } S^{t r}}{r}+\frac{\dot{r} S^{t \phi }}{r}+\frac{\dot{t} S^{r \phi } g'}{2 g}+\frac{\dot{r} S^{t \phi } g'}{2 g} ,\\
\label{S4}
 0 &= &\dot{S^{r \theta }}-\dot{\theta } P^r+\frac{\dot{r} S^{r \theta }}{r}+\frac{g r \dot{\phi } S^{\theta  \phi }}{c^2}+\frac{g \dot{t} S^{t \theta } g'}{2 c^2}-\frac{\dot{r} S^{r \theta } g'}{2 g} ,\\
\label{S5}
 0 &= &\dot{S^{r \phi }}-\dot{\phi } P^r+\frac{\dot{r} S^{r \phi }}{r}+\frac{g \dot{t} S^{t \phi } g'}{2 c^2}-\frac{g r \dot{\theta } S^{\theta  \phi }}{c^2}-\frac{\dot{r} S^{r \phi } g'}{2 g} ,\\
\label{S6}
 0 & = &\dot{S^{\theta  \phi }}+\frac{\dot{\theta } S^{r \phi }}{r}+\frac{2 \dot{r} S^{\theta  \phi }}{r}-\frac{\dot{\phi } S^{r \theta }}{r} .
\eea

The components of equation (\ref{PesKP}) read
\bea
\label{P21}
0 & = & \dot{P^t}-\kappa  P^t+\frac{\dot{t} P^r g'}{2 g}+\frac{\dot{r} P^t g'}{2 g} , \\
\label{P22}
0 &= & \dot{P^r}-\kappa  P^r+\frac{g \dot{t} P^t g'}{2 c^2}-\frac{\dot{r} P^r g'}{2 g}, \\
\label{P23}
0 &= & \dot{P^{\theta }}+\frac{\dot{\theta } P^r}{r}, \\
\label{P24}
0 &= & \dot{P^{\phi }}+\frac{\dot{\phi } P^r}{r} .\\
\eea
Further, the relation $P^\mu U_\mu =0$ reads 
\begin{equation}
  \label{puline}
  g \dot{t} P^t-\frac{c^2 \dot{r} P^r}{g}=0.
\end{equation}
%
\subsection{Deriving the Pseudo-geodesic Radial Solution}

For the scenario (\ref{scenB}), the two constraints are
$V^\mu=\alpha P^\mu$ and $W^\mu=\lambda P^\mu$.
In the radial ansatz, the former reads
\bea
  \label{OC1}
 0 &=& -\frac{c^2 P^r S^{t r}}{g}-\alpha  P^t  ,     \\
\label{OC2}
 0 &= &-g P^t S^{t r}-\alpha  P^r    ,               \\
\label{OC3}
 0 &= &\frac{c^2 P^r S^{r \theta }}{g}-g P^t S^{t \theta}, \\
\label{OC4}
 0 &= &\frac{c^2 P^r S^{r \phi }}{g}-g P^t S^{t \phi },
\eea
and the latter gives
\bea
\label{PL1}
0 &=& \frac{2 c r^2 P^r S^{\theta  \phi }}{g}-2 \lambda  P^t ,\\
\label{PL2}
0 &=& \frac{2 g r^2 P^t S^{\theta  \phi }}{c}-2 \lambda  P^r ,\\
\label{PL3}
0 &=& 2 c P^r S^{t \phi }-2 c P^t S^{r \phi }, \\
\label{PL4}
0 &=& 2 c P^t S^{r \theta }-2 c P^r S^{t \theta } .
\eea
One can solve \eqref{puline} for $P^t$ 
\begin{equation}
  \label{p1pre}
  P^t=\frac{c^2 \dot{r} P^r}{g^2 \dot{t}}.
\end{equation}
Using this result along with equation \eqref{PPCompleta} one gets
\begin{equation}
  \label{RL}
  \dot{r}=\frac{ \pm g   \dot{t}}{c}.
\end{equation}
Thus,
\begin{equation}
  \label{AP1}
  P^t=\frac{ \pm c  P^r}{g}.
\end{equation}
Using \eqref{PL1} along with \eqref{C3} one can solve for $\lambda$ 
\begin{equation}
  \label{L}
  \lambda =-\pm C_3 .
\end{equation}
Now using \eqref{OC1} along with \eqref{AP1} one can solve for $S^{tr}$, giving 
\begin{equation}
  \label{A}
  S^{t r}=-\frac{\pm  \alpha  }{c}.
\end{equation}
Replacing \eqref{AP1} in \eqref{OC3} and using \eqref{C4} one gets
\begin{equation}
  \label{AS13}
  S^{t \theta}=- \frac{ \pm c C_4}{g  r} ,
\end{equation}
and using \eqref{AP1} in \eqref{OC4} with \eqref{C3} gives 
\begin{equation}
  \label{AS14}
  S^{t \phi }=-\frac{\pm c j}{g r} .
\end{equation}
Before restricting the possible angular dependence,
the constants associated to the Killing vectors read
\bea
  \label{C3Completo}
C_3 &=& r \left(r \sin (\phi ) P^{\theta }+r \sin (\theta ) \cos (\theta ) \cos (\phi ) P^{\phi }+\sin (\phi ) S^{r \theta }-r \sin ^2(\theta ) \cos (\phi ) S^{\theta  \phi }+\sin (\theta ) \cos (\theta ) \cos (\phi ) S^{r \phi }\right), \\
  \label{C4Completo}
  C_4 &=& -r \left(r \cos (\phi ) P^{\theta }-r \sin (\theta ) \cos (\theta ) \sin (\phi ) P^{\phi }+r \sin ^2(\theta ) \sin (\phi ) S^{\theta  \phi }+\cos (\phi ) S^{r \theta }-\sin (\theta ) \cos (\theta ) \sin (\phi ) S^{r \phi }\right), \\
  \label{jCompleto}
  j  &=& -r^2 \sin ^2(\theta ) P^{\phi }+r^2 \sin (\theta ) (-\cos (\theta )) S^{\theta  \phi }-r \sin ^2(\theta ) S^{r \phi }.
\eea
Those can be solved for the components of the spin tensor
\bea
  S^{\theta  r}&=&\frac{C_3 \sin (\phi )-C_4 \cos (\phi )-r^2 P^{\theta }}{r}, \\
S^{\theta  \phi }&=&\frac{-C_4 \sin (\phi )-C_3 \cos (\phi )-j \cot (\theta )}{r^2},\\
S^{r \phi }&=&\frac{C_3 \cot (\theta ) \cos (\phi )+C_4 \cot (\theta ) \sin (\phi )+j \cot ^2(\theta )-j \csc ^2(\theta )-r^2 P^{\phi }}{r}.
\eea
Differentiating those three equations
and restricting to the line
 $\theta=\frac{\pi}{2}$ and $\phi=0=P^\theta=P^\phi$ one gets 
\bea
  \label{AS23Punto}
\dot{S^{r \theta }} &=& -r \dot{P^{\theta }}-r \dot{\phi } S^{\theta  \phi }-\frac{\dot{r} S^{r \theta }}{r},\\
\label{AS34Punto}
  \dot{S^{\theta  \phi }} &=& \frac{\dot{\theta } j}{r^2}+\frac{\dot{\phi } S^{r \theta }}{r}-\frac{2 \dot{r} S^{\theta  \phi }}{r}.
\eea
Now, using   \eqref{C3} \eqref{C4}, \eqref{S4}, \eqref{P23}, \eqref{RL}, \eqref{AS13},  and  \eqref{AS23Punto} 
one finally gets 
\begin{equation}
  \label{pphipunto}
  \frac{C_3 \dot{\phi } \left(c^2-g\right)}{c r}=0.
\end{equation}
As \eqref{L} makes $C_3$ proportional to $\lambda$
(where $\lambda$ is the usual nonzero constant associated to helicity in the flat case and flat space should
be a special case of the discussion) concludes that the massless STOP maintains zero angular velocity
$\dot{\phi}=0$.
This result also implies through \eqref{P24} that $\dot{P^\phi}=0$.
Following similar operations with \eqref{P4} using  $\dot{P^\phi}=0$, $\dot{\phi}=0$, \eqref{jConst}, \eqref{C3}, \eqref{C4}, \eqref{RL}, and \eqref{AS14}
one gets
\begin{equation}
  \label{pthetapunto}
  \frac{C_3 \dot{\theta } \left(c^2-g\right)}{c^2 r^2}=0.
\end{equation}
This implies that the massless STOP remains in the
equatorial plane $\dot{\theta }=0$. From \eqref{P23} one also finds $\dot{P^\theta}=0$.
This completes the prove that the solution in the radial direction exists and that it is constant along the trajectory.\\
With this, the complete radial solution is given by
\bea
\label{finalS01}
  S^{t r}        &=&-\frac{\pm \alpha  }{c} , \\
\label{finalS23}
  S^{\theta  \phi }&=&\frac{\pm \lambda }{r^2} , \\
\label{finalS12}
  S^{r \theta}&=&-\frac{C_4}{r} , \\
\label{finalS02}
S^{t \theta}&=&-\frac{\pm c  C_4 }{g r} , \\
\label{finalS03}
S^{t \phi }&=&-\frac{\pm c j  }{g r}  , \\
\label{finalS13}
S^{r \phi }&=&-\frac{j}{r} , \\
\label{finalrt}
\frac{\dot{r}}{\dot{t}}&=&\frac{\pm g }{c}  , \\
\label{finalP0}
P^t&=&\frac{2 c E-\pm  \alpha   g'}{2 c g}  , \\
\label{finalP1}
P^r   &=& \pm\frac{2 c E-\pm \alpha   g'}{2 c^2  } . 
\eea
One sees that indeed $P^2=u^2=0$, just like for massless geodesics without spin.
\subsection{A ``Thermal'' Surprise}

The above solution seems to be almost trivial,
since the STOP travels 
the same lightlike radial path as the spin-less counter part.
However, there is a difference in the energy perceived by an observer at constant $r$.

Let us exemplify this effect by considering a massless STOP heading radially out from a certain radius $r_1$. For this particle one has
\begin{equation}
P^t = \frac{E - \alpha g'/2c}{g}.
\end{equation}
The energy measured by a static observer at radius $r$ (one whose 4-velocity is $U_o^\mu = c(g^{-1/2}(r),0,0,0)$) is then
\begin{equation}
{\mathcal{E}}(r) = U^\mu_o P_\mu =c g^{1/2} P^t = c\frac{E - \alpha g'/2c}{g^{1/2}}.
\end{equation}
Using that $E$ is a constant of motion, we can relate the energy measured at two radii $r_1$ and $r_2$ by
\begin{equation}
g^{1/2}(r_1) {\mathcal{E}}(r_1) + \frac{\alpha}{2}g'(r_1) = g^{1/2}(r_2) 
{\mathcal{E}}(r_2) + \frac{\alpha}{2}g'(r_2)
\end{equation}
so
\begin{equation}
{\mathcal{E}}(r_2) = \sqrt{\frac{g(r_1)}{g(r_2)}}{\mathcal{E}}(r_1) + \frac{\alpha}{2}\frac{g'(r_1) - g'(r_2)}{g^{1/2}(r_2)}.
\end{equation}
Notice that if $\alpha = 0$, the usual gravitational redshift formula is recovered. If $\alpha \neq 0$, a new effect appears, namely, an extra contribution to the measured energy due to the spin. In particular, if we consider a Schwarzschild black hole, and 
that the particle is emitted right outside the event horizon, the above formula predicts a \emph{non-vanishing} energy measured at infinity ($r_2 \longrightarrow \infty$), given by
\begin{equation} 
{\mathcal{E}}_\infty = \frac{\alpha c}{2 r_s }= \frac{2\pi\alpha}{\hbar} k_B T_H, 
\end{equation}
where $k_B$ is the Boltzmann constant and $T_H$ is the Hawking temperature.
Like in the spin-less case, any emission of finite energy 
from the close vicinity of the black hole horizon
experiences a red shift when propagating towards radial infinity. However, there are two remarkable differences
\begin{itemize}
\item
While in the spin-less case this
redshift actually leaves no energy at all at radial infinity, the STOP will have some finite energy ${\mathcal{E}}_\infty$ at
radial infinity.
\item
The amount of ${\mathcal{E}}_\infty$ is determined by the surface gravity of the black hole
background 
\be\label{Hawk}
 {\mathcal{E}}_{\infty}  = \frac{\alpha}{c}\cdot \frac{1}{2}g'|_{r=r_s},
\ee
which is identical to the Hawking temperature \cite{Hawking:1974rv},
if one chooses  $\alpha=\hbar /(2\pi)$.
\end{itemize}
Given the fact that the classical dynamics of a
massless STOP has no
obvious conceptual connection with the spin-independent quantum effects of black hole thermodynamics,
the appearance of the Hawking relation (\ref{Hawk})
is quite surprising.

\section{Conclusions}
\label{secConcl}

In this paper, the possibility of non-geodesic motion of
massless STOPs is revisited. It is found that, in contrast to the common belief, a consistent and nontrivial formulation
of massless STOPs within the equations (\ref{momentummotion}, \ref{spinmotion})
is actually possible.
This possibility arises by analyzing various constraints, 
which have not been previously considered (summarized in Table \ref{tabcond}).
The constancy of those constraints is shown.
Finally, the integration of the equations (\ref{momentummotion}, \ref{spinmotion}) 
combined with the new constraints is discussed in light of a simple example.
By studying spherically symmetric background spacetimes which fulfill
the condition $g_{00}=-c^2/g_{11}$, a nontrivial solution is obtained
for purely radial motion (\ref{finalS01}-\ref{finalP1}).
This solution is then discussed for the radial motion of massless STOPs
which are produced with finite energy at the close vicinity of the black hole horizon.
It is found that the energy of the massless STOPs at radial infinity is given by
the spin parameter times the surface gravity of the background horizon (\ref{Hawk}).
This is the same metric dependency as it appears in the Hawking relation.

\section*{Acknowledgments}
We thank F. Asenjo and I. A. Reyes for discussions and suggestions. 
The work of B.K. and C.A. was supported by Fondecyt Project 1120360 and Anillo Atlas Andino 10201. 
The work of N.Z. was supported by CONICyT-Chile grant No 21080567.
\newpage
\begin{appendix}

\section{Equations of Motion from a Lagrangian Formulation}\label{EOMfromL}

The equations of motion can be obtained by considering variations of the action with respect to the independent
variations $\delta x^\mu$ and
$\delta \theta^{\mu \nu} \equiv \eta^{a b}
{e_{a}}^{\mu}({\delta e_{b} }^\nu + \Gamma^{\nu}_{\lambda \rho} {e_{b}}^{\lambda}\delta x^{\rho}) = - \delta
\theta^{\nu \mu}$. 
It is important to note that one has to vary with respect to 
$\delta \theta^{\mu \nu}$ and not with respect to ${e_{a}}^\mu$.
If arbitrary variations in ${e_{a}}^\mu$ were used,
one would consider too many degrees of freedom.
This problem is avoided
by the use the variation $\delta \theta^{\mu \nu}$, which has only
six degrees of freedom.
Before proceeding with the variation of the Lagrangian,
let us derive a 
relation between the variation of the angular velocity, $\delta \sigma^{\mu\nu}$, and $\delta\theta^{\mu\nu}$. 
For this, one defines the 
following symbols for covariant derivative and covariant variation:
\begin{align}
\mathring{A^\mu}&\equiv \dot{A^{\mu}}+{{\Gamma}^{\mu}}_{\lambda \rho}A^{\lambda} u^{\rho},\\
D{A^\mu}&\equiv \delta A^{\mu}+{\Gamma^{\mu}}_{\lambda \rho}A^{\lambda} \delta x^{\rho}.
\end{align}
Similarly, one has
\begin{equation}
D\sigma^{\mu \nu} \equiv \delta \sigma^{\mu \nu} + {\Gamma^\mu}_{\alpha \beta}\sigma^{\beta \nu}\delta x^\alpha
+{\Gamma^\nu}_{\alpha \beta}\sigma^{\mu \beta }\delta x^\alpha.
\end{equation}
Solving this for $\delta \sigma^{\mu \nu}$ and
using the relations
\begin{equation}
  \label{eqSigmaEtaDef}
  D\sigma^{\mu\nu}-\mathring{(\delta\theta^{\mu\nu})}=D(\eta^{a b}{e_{a}}^{\mu}\mathring{{e_{b}}^{\nu}})-\mathring{(\eta^{a b}{e_{a}}^{\mu}D{e_{b}}^{\nu})}
\end{equation} 
and
  \begin{equation}
    \label{opCom}
    D(\mathring{A^{\mu}})-\mathring{(D A^{\mu})}={R^{\mu}}_{\lambda \alpha \beta}A^{\lambda}u^{\beta}\delta x^{\alpha}
  \end{equation}
along with the definition of $\sigma^{\mu\nu}$, one can express the variation $\delta \sigma^{\mu \nu}$
in terms of the variations $\delta x^\mu$ and $\delta \theta^{\mu \nu}$ as
\begin{equation}
  \label{eqSigmaTetaFinal}
  \delta \sigma^{\mu \nu}=\mathring{(\delta \theta^{\mu \nu})}+{\sigma_\alpha}^\nu \delta \theta^{\alpha \mu}-{\sigma_\alpha}^\mu \delta \theta^{\alpha \nu}+(g^{\mu \lambda}{R^{\nu}}_{\lambda \beta \alpha}u^{\alpha}-{\Gamma^{\mu}}_{\lambda \beta}\sigma^{\lambda\nu} -{\Gamma^{\nu}}_{\lambda \beta}\sigma^{\mu \lambda})\delta x^\beta.
\end{equation}
With this result at hand one can proceed with
the variation of the Lagrangian with respect to $\delta x^\alpha$, which is given by
\begin{equation}
  \label{eqQDef}
  \frac{\partial L}{\partial x^\alpha}\delta x^\alpha =\frac{\partial L}{\partial g_{\mu\nu}}g_{\mu \nu , \alpha}\delta x^\alpha,
\end{equation}
where
\begin{equation}\label{eqQDefEx1}
\frac{\partial L}{\partial g_{\mu\nu}} = U^\mu U^\nu L_1 + 2 \sigma^{\mu \rho} {\sigma_{\rho}}^\nu L_2 + (U^\mu \sigma^{\nu\lambda}\sigma_{\lambda\rho}U^\rho + U_\lambda \sigma^{\lambda \mu}\sigma^{\nu \rho}U_\rho + U^\lambda \sigma_{\lambda\rho}\sigma^{\rho\mu}U^\nu)L_3 + 4\sigma^{\mu\lambda}\sigma_{\lambda\rho}\sigma^{\rho\alpha}{\sigma_\alpha}^\nu L_4.
\end{equation}
Using \eqref{PDef} and \eqref{SDef} one 
can write $\partial L/\partial g_{\mu\nu}$ as
\begin{equation}
  \label{eqQRes}
 \frac{\partial L}{\partial g_{\mu\nu}}= - \frac{1}{4}(P^\mu U^\nu + P^\nu U^\mu)+\frac{1}{4}(S^{\mu \alpha}{\sigma_\alpha}^\nu+S^{\nu \alpha}{\sigma_\alpha}^\mu).
\end{equation}
By invoking \eqref{PDef} and \eqref{SDef} again, 
one can further show that
\begin{equation}
  \label{eqRelPUSS}
  P^\mu U^\nu -P^\nu U^\mu =  S^{\mu \alpha}{\sigma_\alpha}^\nu-S^{\nu \alpha}{\sigma_\alpha}^\mu.
\end{equation}
Using the above relations, one can get the following 
variation of the action:
\begin{align}
  \nonumber
  \delta I &= \int d\tau \left\{\frac{\partial L}{\partial U^\mu}\delta U^\mu + \frac{1}{2}\frac{\partial L}{\partial \sigma^{\mu\nu}}\delta\sigma^{\mu\nu} + \frac{\partial L}{\partial x^\mu}\delta x^\mu\right\}\\\nonumber
&= \int d\tau \left\{ -P_\mu (\delta \dot{x}^\mu+{\Gamma^{\mu}}_{\nu \beta}U^\nu \delta x^\beta) - \frac{1}{2}S_{\mu\nu}g^{\mu\lambda}{R^{\nu}}_{\lambda \beta \alpha}U^\alpha \delta x^\beta - 
  \frac{1}{2}S_{\mu\nu}\left[\mathring{(\delta \theta^{\mu\nu})}+{\sigma_\alpha}^\nu \delta \theta^{\alpha\mu}-{\sigma_\alpha}^\mu \delta \theta^{\alpha\nu}\right] \right\}\\ 
&= \int d\tau \left\{ \left(\mathring{P_\beta} + \frac{1}{2}R_{\beta \alpha \lambda \nu}U^\alpha S^{\lambda\nu}\right)\delta x^\beta + \frac{1}{2}(\mathring{S_{\mu\nu}} + S_{\nu\alpha}{\sigma^\alpha}_\mu - S_{\mu\alpha}{\sigma^\alpha}_\nu)\delta \theta^{\mu\nu} - \dot{(P_\mu \delta x^\mu)} - \frac{1}{2}\dot{(S_{\mu\nu}\delta \theta^{\mu\nu})} \right\}.
\label{varFinal2}
\end{align}
Finally, imposing a vanishing variation (\ref{varFinal2}) 
implies the well known equations of motion
\begin{align}
 \frac{D P^\mu}{D\tau}&=-\frac{1}{2}{R^\mu}_{\nu\alpha\beta}U^\nu S^{\alpha\beta},\\ 
\frac{D S^{\mu \nu}}{D\tau}&=S^{\mu
\lambda}{\sigma_\lambda}^\nu-\sigma^{\mu
\lambda}{S_\lambda}^\nu=P^\mu U^\nu-U^\mu P^\nu.
\end{align}


\section{Constant of Motion Associated to Killing Vector}
\label{SecNoether}

Consider the general variation of the Lagrangian found in Appendix \ref{EOMfromL},
\begin{equation}\label{generalvariation}
\delta L =  \left(\frac{DP_\beta}{D\tau} + \frac{1}{2}R_{\beta \alpha \lambda \nu}U^\alpha S^{\lambda\nu}\right)\delta x^\beta  + \frac{1}{2}\left(\frac{DS_{\mu\nu}}{D\tau} + S_{\nu\alpha}{\sigma^\alpha}_\mu - S_{\mu\alpha}{\sigma^\alpha}_\nu\right)\delta \theta^{\mu\nu} - B,
\end{equation}
where
\begin{equation}
B \equiv \frac{d}{d\tau}\left(P_\mu \delta x^\mu + \frac{1}{2}S_{\mu\nu}\delta\theta^{\mu\nu}\right).
\end{equation}
Also, let $\xi^\mu$ be a Killing vector of a given metric $g_{\mu\nu}$, 
\begin{equation}
\mathcal L_\xi g_{\mu\nu} = \xi_{\mu;\nu} + \xi_{\nu;\mu} = 0,\label{killingequation}
\end{equation}
where $\mathcal L_\xi$ denotes Lie derivative along the vector $\xi^\mu$. We shall prove that the particular transformation 
\begin{equation}\label{symmetrytransformation}
\delta_\xi x^\mu \equiv \xi^\mu;\qquad \delta_\xi \theta^{\mu\nu} \equiv -g^{\nu\alpha}{\xi^{\mu}}_{;\alpha} = -\xi^{\mu;\nu}
\end{equation}
is a Noetherian symmetry, i.e. $\delta_\xi L$ is a total derivative for \emph{any} trajectory, \emph{without using the equations of motion}\footnote{For a recent review on Noether's Theorem and Noetherian symmetries, see \cite{banados16}.}. The conserved charge associated to this symmetry is
\begin{equation}
Q_\xi = P_\mu \xi^\mu - \frac{1}{2}S^{\mu\nu}\xi_{\mu;\nu}.
\end{equation}
First, recalling the general variation $\delta\theta^{\mu\nu}$ given in Appendix \ref{EOMfromL}, we point out that
\begin{align}\nonumber
\delta_\xi\theta^{\mu\nu} &= \eta^{ab}{e_{a}}^\mu\,\delta_\xi {e_{b}}^\nu + \eta^{ab}{e_{a}}^\mu\,{\Gamma^\nu}_{\lambda\beta}\,{e_{b}}^\lambda\,\xi^\beta.\\\nonumber
&= \eta^{ab}{e_{a}}^\mu\,\delta_\xi {e_{b}}^\nu + g^{\mu\lambda}{\xi^\nu}_{;\lambda} - {g^\mu\lambda}{\xi^\nu}_{,\lambda}\\
&= \eta^{ab}{e_a}^\mu\,\mathcal L_\xi{e_b}^\nu - \xi^{\mu;\nu},
\end{align}
where $\mathcal L_\xi{e_b}^\nu = \delta_\xi {e_{b}}^\nu - {e_b}^\lambda\,{\xi^\nu}_{,\lambda}$, with $\delta_\xi {e_{b}}^\nu = {{e_b}^\nu}_{,\alpha} \xi^\alpha$. In the second equality above, we used the definition of covariant derivative and the relation between the tetrads and the metric. Imposing $\mathcal L_\xi{e_b}^\nu = 0$ ($\xi^\mu$ is a Killing vector), we explain here the motivation of the transformation for $\delta_\xi\theta^{\mu\nu}$. Now, in Eq. \eqref{generalvariation}, the third term is already written as a total derivative, so we pay attention on the other two terms. Using an integration by parts, the first term of \eqref{generalvariation} can be rewritten as
\begin{equation}
\binom{\text{first}}{\text{term}} = \frac{d}{d\tau}\left(P_\mu \xi^\mu\right) - P_\mu {\xi^\mu}_{;\nu}U^\nu - \frac{1}{2}S^{\beta\nu}R_{\beta\nu\rho\lambda}\xi^\lambda U^\rho,\label{firstterm}
\end{equation}
Using the known identity for a Killing vector $\xi_{\alpha;\beta;\mu} = R_{\alpha\beta\mu\rho}\xi^\rho$ and decomposing $P^\mu U^\nu$ into its symmetric and antisymmetric part, one sees that
\begin{equation}
\binom{\text{first}}{\text{term}} = \frac{d}{d\tau}\left(P_\mu \xi^\mu\right) - \frac{1}{2}\xi_{\mu;\nu}\left(P^\mu U^\nu - U^\mu P^\nu\right) - \frac{1}{2}S^{\mu\nu}\frac{D}{D\tau}\left(\xi_{\mu;\nu}\right).
\end{equation}
In the second term of Eq. \eqref{generalvariation}, we can replace the identity of Eq. \eqref{eqRelPUSS}. After putting all this together and collecting terms, the variation of the Lagrangian reads
\begin{equation}
\delta_\xi L = \frac{d}{d\tau}\left(P_\mu \xi^\mu\right) - \frac{1}{2}\left(P_\mu U_\nu - U_\mu P_\nu\right)\left(\xi^{\mu;\nu} + \delta_\xi\theta^{\mu\nu}\right) - \frac{1}{2}S^{\mu\nu}\frac{D}{D\tau}\left(\xi_{\mu;\nu}\right) + \frac{DS_{\mu\nu}}{D\tau}\delta_\xi\theta^{\mu\nu} - B.
\end{equation}
so, for $\delta_\xi\theta^{\mu\nu} = -\xi^{\mu;\nu}$, $\delta_\xi L$ clearly vanishes ($B$ is of course evaluated using Eqs. \eqref{symmetrytransformation} as well). This is equivalent to say that $\delta_\xi L$ is the total derivative of a constant number, which without loss of generality, we can set equal to zero. On the other hand, if the equations of motion hold, a generic variation of the Lagrangian reads
\begin{equation}
\delta_{os}L = \frac{d}{d\tau}\left(P_\mu \delta x^\mu + \frac{1}{2}S_{\mu\nu}\delta\theta^{\mu\nu}\right).
\end{equation}
In particular, the latter must hold for the variations defining the symmetry above, so comparing both variations, we get the conservation law
\begin{equation}
0= \frac{d}{d\tau}\left(P_\mu \xi^\mu - \frac{1}{2}S^{\mu\nu}\xi_{\mu;\nu} \right),
\end{equation}
and thus $Q_\xi$ is a constant of motion.

\section{Auxiliary Results}
\subsection{Parallel Lightlike Vectors}
\label{Atheorem}

In this section the following useful relation will be proven:
for a vector $A^\mu$ and a time or lightlike vector $B^\mu$ (both non-vanishing)
in a spacetime equipped with an invertible metric it is true that
\be\label{Etheorem}
\mbox{if}\;
\left\{
\begin{array}{cc}
  A^\mu A_\mu=0& \,\&\\
    A^\mu B_\mu=0& \, \&\\
    B^\mu B_\mu\ge 0
\end{array}
\right\}\; \mbox{then}\;\Rightarrow  B^\mu=\kappa A^\mu.
\ee

For the proof let us consider the
invertible spacetime tetrad ${e_{a}}^{\mu}(x)$ with inverse ${e^{a}}_{\mu}(x)$ (not to be confused with the spin tetrad defined throughout this work),
defined by the equations
\bea
   {e_{a}}^{\upsilon}{e_{b}}^{\nu}g_{\mu\nu}&= & \eta_{ab},\\ \nonumber
   {e_{a}}^{\mu}{e_{b}}^{\nu} \eta^{ab}&= & g^{\mu\nu}.
\eea
With this at hand, we can map four-vectors by means of the transformation
\begin{equation}
  \label{AppendixmapTangent}
   V_a= {e_{a}}^{\mu}V_\mu,
    \end{equation}
which clearly leaves the norm of a four-vector invariant when the inverse transformation is considered.
So we must only prove the desired result in Minkowski space. 
To begin, let us write $A^a$ as
\begin{equation}
  A^a=(a^0, a^1, a^2, a^3).
\end{equation}
Without loss of generality, we can rotate our coordinates such that $a^2=a^3 = 0$. 
Using that $A^a$ is lightlike, we get
\begin{equation}
  a^0 = \pm a^1.
\end{equation}
Writing $B^b=(b^0, b^1, b^2,b^3)$ and using the orthogonality of $A^a$ and $B^b$, we find that
\begin{equation}
  0 = a^1(\pm b^0 - b^1),
\end{equation}
from which we conclude that  $b^1=\pm b^0$. This implies that
\begin{equation}
  \label{acasicasi}
 B_a B^a = -(b^2)^2 - (b^3)^2.
\end{equation}
This implies  $b^2=b^3=0$, since $B_a B^a  \geq 0$.
With this result we get the final relation between $A^a$ and $B^b$
\begin{equation}
  \label{appfinal}
  B_a=\frac{b_0}{a_0} A_a \propto A_a,
\end{equation}
which we can map back to coordinate spacetime, finishing the proof.
\subsection{Relations Between Spin and Proportionality Constants}
\label{SecJ2rel}

At some points of the discussion also the following three relations will be used.
Let $S^{\mu\nu}$ be an antisymmetric four-tensor. Contracting the identity \cite{hanson74}
\begin{equation}\label{id1}
{S^{*\mu}}_\alpha S^{\alpha\nu}= {S^\mu}_\alpha S^{*\alpha\nu} = - \frac{1}{4}g^{\mu\nu} S^{\alpha\beta}{S^{*}}_{\alpha\beta},
\end{equation}
with $P_\mu P_\nu$, one gets the condition
\begin{equation}
W_\mu V^\mu = \frac{1}{4}P_\mu P^\mu S^{\alpha\beta} {S^*}_{\alpha\beta}.
\end{equation}
It remains to verify the relations between $J^2$ and $\lambda^2$ listed in (\ref{scenA})-${\mathcal A}$
and in (\ref{scenB})-${\mathcal B}$.
As said, the constant $\lambda$ is what is usually considered the helicity, 
while $\alpha$ is a new constant introduced. The relation
between both can be obtained from contracting $W^\mu$ with $S_{\beta\mu}$ and using Eqs
\eqref{id1}. The result is
\begin{equation}\label{SSstar}
- \frac{1}{4}S^{*\alpha\beta}S_{\alpha\beta} P_\mu = \lambda \alpha P_\mu.
\end{equation}
We conclude that
\begin{equation}
\alpha = - \frac{S^{*\alpha\beta}S_{\alpha\beta}}{4\lambda}.\label{alpha}
\end{equation}
At this point it should be pointed out that if we demand 
\begin{equation}
S^{*\alpha\beta}S_{\alpha\beta} = 0,
\end{equation}
then $\alpha = 0$. On othe other hand, 
recalling that $\alpha$ is an eigenvalue of ${S^{\mu}}_\nu$ with eigenvector $P^\mu$,
we have the characteristic equation
\begin{equation}
\label{eigAlpha}
\alpha^4 + \alpha^2S^2 - \frac{1}{16}(S^{*\alpha\beta}S_{\alpha\beta})^2 = 0.
\end{equation}
Combining \eqref{alpha} in the latter, we arrive at the new equation
\begin{equation}
\lambda^4 - \lambda^2S^2 - \frac{1}{16}(S^{*\alpha\beta}S_{\alpha\beta})^2 = 0.\label{charlambda}
\end{equation}
Eq.~\eqref{charlambda} coincides with the characteristic equation for ${S^{*\mu}}_\nu$,
regarding $\lambda$ as its eigenvalues. 
Substracting \eqref{charlambda} from \eqref{eigAlpha} we get:
 \begin{equation}
   \label{eigMix}
   \alpha^4-\lambda^4+S^2(\alpha^2+ \lambda^2)=(\alpha^2+\lambda^2)(\alpha^2-\lambda^2+S^2)=0.
 \end{equation}
From this, we can conclude that, if we demand the use of only real values for $\alpha$ and $\lambda$, the relation 
$J^2=\lambda^2-\alpha^2$ always holds.

\end{appendix}

\end{document}